\documentclass[twocolumn, longbibliography,floatfix]{revtex4-1}

\usepackage{soul}
\usepackage{amsmath}   
\usepackage{amssymb}
\usepackage{graphicx}
\usepackage{dcolumn}
\usepackage{bm}
\usepackage{amsmath}
\usepackage{graphicx}
\usepackage{amsfonts}
\usepackage{subfigure}
\usepackage{graphicx}
\usepackage{array}
\usepackage{float}
\usepackage{color}
\usepackage{amssymb}%
\usepackage[colorlinks=true,linkcolor=blue]{hyperref}%
\hypersetup{allcolors=blue}
\usepackage[normalem]{ulem}
\usepackage{xcolor}

\newcommand{\ket}[1]{\ensuremath{\left\vert #1 \right\rangle}}
\usepackage{mathtools}
\DeclarePairedDelimiterX\braket[2]{\langle}{\rangle}{#1 \delimsize\vert #2}

\begin{document}

\title{Electric field analysis in a cold-ion source using Stark spectroscopy of Rydberg atoms}

\date{\today }

\author{Alisher Duspayev}
    \email{alisherd@umich.edu}
\affiliation{Department of Physics, University of Michigan, Ann Arbor, MI 48109, USA}
\author{Georg Raithel}
\affiliation{Department of Physics, University of Michigan, Ann Arbor, MI 48109, USA}

\begin{abstract}
We analyze electric fields in ion sources 
generated by quasi-continuous photo-ionization of cold Rb atoms trapped in the focal spot of a near-concentric, in-vacuum cavity for 1064-nm laser light. Ion streams are extracted with an external electric field, ${\bf{F}}$. Stark effects of Rb 57$F$
and of nearby high-angular-momentum Rydberg levels, which exhibit large, linear Stark shifts, are employed to study the net electric-field probability distribution within the ion-source region over an extraction-field range of $0<F<0.35$~V/cm. For $F=0$, we also investigate ion-field-induced Stark spectra of the 60$P_{1/2}$-state, which exhibits a (lesser) quadratic electric-field response that affords a simplified electric-field analysis. Experimental Rydberg spectra are compared with theoretical Stark spectra, which are weighed with net electric-field distributions obtained from classical ion-trajectory simulations that include Coulomb interactions. Experiments and models agree well. At small $F$ and high ion source rates, the field approximately follows a Holtsmark distribution, and the ion streams are degraded by the Coulomb micro-fields. With increasing $F$ and at lower ion source rates, the fields become narrowly distributed around ${\bf{F}}$, resulting in directional ion streams that are less degraded by micro-fields. Our results are of interest for monitoring cold-ion sources for focused-ion-beam applications, where Coulomb interactions are of concern, and for studies of electric fields in cold plasmas.
\end{abstract}

\maketitle

\section{Introduction}
\label{sec:intro}

The use of atomic probes and related methods in laser-generated, small-laboratory plasmas, both in cold-atom environments~\cite{feldbaum, abel2011, Bounds_2019, Viray2020} as well as in room-temperature atomic vapors~\cite{weller2016, anderson2017, weller2019, Ma20}, is an area of research of considerable current interest. 
Atom-based electric-field sensing has practical applications in electromagnetic-field metrology~\cite{Sedlacek2012, Holloway2014, degenrmp, anderson2021, RFMS} and quantum control~\cite{idziaszek, secker2016, engel, wang}. 
Recent work on ion plasmas and ion imaging includes ion microscopy~\cite{Schwarzkopf2013, veit2021}, association of molecular ions~\cite{duspayev2021, deiss2021, zuber2022, weckesser, mohammadi}, coupling to ultracold plasmas~\cite{robinson2000, morrison, crockett, weller2019} as well as Rydberg spectroscopy in the presence of ions~\cite{Viteau11, ewald}.
Atom-ion interactions have further attracted interest in quantum chemistry at ultracold temperatures~\cite{coldchembook,  da_Silva_Jr_2015, weckesser, hirzlerprl2022}, many-body dynamics~\cite{meir, dieterle, feldker}, precision measurements~\cite{tacconi2011, loh2013, germann2014, cairncross, benschlomi} and emerging technologies for quantum computing~\cite{doerk, ladd2010, Saffman_2016} and simulation~\cite{blatt2012, bloch2012, bissbort}. While methods to harness such interactions are being investigated in the aforementioned applications, detrimental effects caused by them are also being studied~\cite{schmid, zipkesnature, zipkesprl, casteels, cetina2012}. 

In industrial applications, laser-cooled atoms are employed as a source of focused ion beams (FIBs). Configurations based on magneto-optical trapping~\cite{claessens, hanssen} and on atomic beams cooled in two transverse directions~\cite{Steele_2017, viteau, shayeganrad} have been demonstrated. These approaches present feasible alternatives to other FIB sources that include liquid-metal~\cite{bischoff, gierak} and gas-field~\cite{ward2006} ion sources as well as inductively coupled plasma sources~\cite{smith2006}. Inter-particle Coulomb interactions remain a challenging aspect in cold-atom FIB sources~\cite{knuffman2013}, their applications in industry~\cite{bassim2014recent}, as well as in fundamental science~\cite{emma2010first,schreiber_faatz_2015}.  

A noninvasive, integrated, in-situ atomic electric-field measurement method can be valuable to control Coulomb effects in cold-ion sources. To that end, in the present work we investigate the electric fields in
ion streams prepared by quasicontinuous laser ionization of cylindrical samples of laser-cooled and -trapped Rb atoms. The samples are prepared in the focal region of a far-off-resonant optical-lattice dipole trap (OLDT) that is formed inside a near-concentric, in-vacuum resonator. Noninvasive electric-field measurement is performed 
by laser spectroscopy of the Stark effect of low- and high-angular-momentum Rydberg atoms. Spectra are taken over a range of amplitudes of an applied ion extraction field, ${\bf{F}}$. We explore how the electric-field distribution in the ion-sourcing region transitions from a microfield-dominated distribution at $F=0$~V/cm, which approximates a Holtsmark distribution~\cite{holtsmark}, into a relatively narrow distribution at large enough $F$. 
The Rydberg spectra reflect how the field $F$ turns a Coulomb-pressure-driven, widely dispersed and largely isotropic ion stream into a directed ion beam with reduced Coulomb interactions.

The paper is organized as follows. The utilized experimental setup and methods for creating ion streams and for measuring the electric fields in them are described in Sec.~\ref{sec:setup}. The results of two sets of experiments with rubidium Rydberg $57F$ and $60P$ states are presented and analyzed in Sec.~\ref{sec:results}. The findings are discussed in a broader context in Sec.~\ref{sec:disc}, and the paper is concluded in Sec.~\ref{sec:conc}.

\section{Experimental methods}
\label{sec:setup}

\subsection{Overview}

\begin{figure}[htb]
 \centering
  \includegraphics[width=0.48\textwidth]{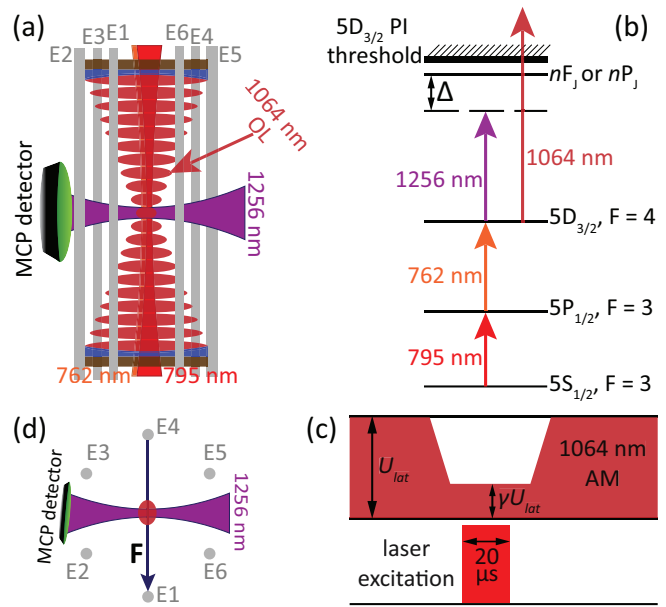}
  \caption{(a) Sketch of the experimental setup and configuration of laser beams. The 1064-nm, 795-nm and 762-nm beams are coupled to the in-vacuum OLDT cavity from below, while the 1256-nm beam is incident from the side. 6 electrodes surrounding the cavity (E1 through E6) are used for electric field control and field ionization (FI) of Rydberg atoms. Rydberg electrons from FI are directed by means of the electrodes to a microchannel plate (MCP) detector for electron counting and data acquisition. (b) Diagram of the utilized $^{85}$Rb energy levels (not to scale). (c) Timing of OLDT amplitude reduction and laser excitation pulses. The small residual OLDT power of 1064~nm light ($\gamma \lesssim 5 \times 10^{-3}$) during laser excitation photoionizes $5D_{3/2}$ atoms. The ion sourcing generates a quasicontinuous ion stream, which is driven by Coulomb repulsion and an external extraction electric field, ${\bf{F}}$. The electric fields in the ion stream are analyzed by Rydberg Stark spectroscopy. (d) Top view of the setup in (a), showing the orientation of $\textbf{F}$.}
  \label{fig1}
\end{figure}

To prepare and probe ion streams, a cylindrically symmetric, long and thin atom cloud is prepared in an OLDT focus, as shown in Fig~\ref{fig1}. With the OLDT adiabatically lowered to less than 1$\%$ of its original depth, photoionization (PI) and Rydberg-atom excitation lasers are simultaneously turned on, creating a quasicontinuous situation in which ions are sourced at a controllable rate within the small OLDT region. The ions stream outwards due to Coulomb repulsion. Under absence of external electric fields, the ensuing quasicontinuous ion stream is cylindrically symmetric around the OLDT axis. When we apply the external dc ion extraction field, ${\bf{F}}$, the ions are extracted and form a continuous, directed ion stream with a reduced ion density and diminished effects from Coulomb repulsion. The ions in the quasisteady-state streams are subject to a net electric field, ${\bf{E}}_{net}$, which consists of ${\bf{F}}$, a macroscopic, smoothly varying field caused by the average charge density distribution, ${\bf{E}}_{mac}$, and a fluctuating microscopic field ${\bf{E}}_{mic}$ caused by the discreteness of the ionic charges, such that
\begin{equation}
{\bf{E}}_{net} = {\bf{F}} + {\bf{E}}_{mac} + {\bf{E}}_{mic}.
\label{eq:e_net} 
\end{equation}
\noindent In our experiment, Rydberg atoms are excited concurrently with the ion sourcing. The objective of our study is to use Rydberg-atom Stark effects to measure the distribution of the field-magnitude, $\vert {\bf{E}}_{net} \vert$, as a function of the ion-source rate, $R_{ion}$, and the magnitude of the extraction field, $F$. 
Our experimental setup shares some aspects with other hybrid  systems~\cite{harter, tomza} of cold atoms and ions. Here, we aim at a spectroscopic measurement of electric fields in ion sources, and there is no ion trap involved.

\subsection{Experimental setup}

Our experimental setup is sketched in Fig.~\ref{fig1}~(a). Cold $^{85}$Rb atoms are loaded from a 3D magneto-optical trap (MOT) into the OLDT of depth $U_{latt} \sim h \times 8$~MHz for 5$S_{1/2}$-atoms, generated by coupling 1064~nm light into a clean TEM$_{00}$ mode of an in-vacuum, vertically oriented, nearly concentric optical cavity of finesse $\sim 600$.
Up to $\sim 10^4$ atoms are collected in the high-intensity regions of the OLDT cavity mode while the MOT is on, with cooling provided by the MOT laser field~\cite{Chen2014praatomtrapping}. The OLDT provides atom samples with a density $\lesssim 10^{11}$~cm$^{-3}$, a diameter of $\sim 20~\mu$m, and a length of $\lesssim 1$~mm along the OLDT axis. Details on cavity stabilization and atom preparation are provided in~\cite{Chen2014praatomtrapping}. After loading, $U_{latt}$ is adiabatically ramped down to a reduced depth of $\gamma U_{latt}$ by means of an acousto-optical modulator 
for controlled three-photon PI of atoms trapped in the OLDT, allowing us to set the ion-source rate, $R_{ion}$.

Four lasers, including the OLDT laser, are used to generate ion streams via PI and to measure ion fields via Rydberg-atom spectroscopy, as shown in the $^{85}$Rb-level scheme in Fig.~\ref{fig1}~(b). The timing of OLDT amplitude reduction and laser excitation is depicted in Fig.~\ref{fig1}~(c). The MOT beams are turned off during ionization and Rydberg-atom excitation,
while the MOT repumper beam is left on. The two probe lasers that drive the lower $\ket{5S_{1/2}, F = 3} \rightarrow \ket{5P_{1/2}, F = 3}$ (795~nm, power $\sim 200$~nW before vacuum chamber) and middle $\ket{5P_{1/2}, F = 3} \rightarrow \ket{5D_{3/2}, F = 4}$ (762~nm, power $\sim 1~\mu$W before vacuum chamber) transitions are pulsed on for 20~$\mu$s. The lasers are on-resonance with the respective transitions, and the laser beams are coupled through the cavity which ensures good spatial overlap with the atoms trapped in the utilized TEM$_{00}$ mode of the OLDT. More details on the laser control are provided in~\cite{cardman2021}. 
The laser driving the $\ket{5D_{3/2}, F = 4} \rightarrow \ket{\, Rydberg \, }$ transitions (1256~nm, power $\sim 15$~mW before chamber) is always on, and its detuning, $\Delta$, is scanned across the Rydberg states of interest. The 1256~nm laser is introduced from a direction perpendicular to the OLDT axis, as seen in Figs.~\ref{fig1}~(a) and~(d).
A cylindrical lens installed outside the vacuum system is used to shape the 1256~nm laser mode to match the OLDT waist of $\approx$ 20~$\mu$m and to cover the length of the trapped-atom cloud along the OLDT axis ($\lesssim 1$~mm). The frequency scans of the 1256~nm laser are calibrated by sending a beam sample through a Fabry-P\'erot etalon (374~MHz free spectral range), the transmission peaks of which provide frequency marks.

The OLDT cavity is surrounded by six long, thin electrodes parallel to the OLDT axis (E1 through E6 in Fig.~\ref{fig1}). The electrodes are used to apply electric fields during ion sourcing and Rydberg-atom excitation, as well as to field-ionize the Rydberg atoms for Rydberg-atom counting after the 20~$\mu$s probe duration.   
During the 20~$\mu$s ion-sourcing and Rydberg probing, the electrodes E2, E3, E5 and E6 are kept at voltages that minimize dc Stark shifts of Rydberg levels, while voltages applied to the electrodes E1 and E4 are employed to apply the controllable, approximately homogeneous 
ion extraction field, ${\bf{F}}$, within the experimental region
[see Fig.~\ref{fig1}~(d)]. The field ${\bf{F}}$, when applied, directs the ion flow in a direction transverse to the OLDT axis. The 
electrode arrangement allows electric-field control in the two directions transverse to the OLDT axis. In previous work using the same setup~\cite{Chen2014praatomtrapping, jamie, cardman2021} there has been no indication of stray electric-field components parallel to the OLDT axis. For field ionization (FI) of Rydberg atoms, we apply individually controlled high-voltage pulses 
to electrodes E5 and E6. In contrast to  previous experiments~\cite{Chen2014praatomtrapping, jamie, cardman2021, atoms10040117}, where positive-ion detection was used, here we detect Rydberg electrons liberated by FI. This modification is necessary in order to discern the spectroscopic Rydberg-atom signal from the ions generated 
by PI of 5$D_{3/2}$ atoms. 
Small control-voltage adjustments applied to E2 and E3 steer the electrons onto a microchannel plate detector (MCP). MCP pulses are counted and recorded for processing. Rydberg spectra are obtained as a function of laser detuning, $\Delta$, the magnitude $F$, and the ion-source rate, $R_{ion}$.

After loading atoms into the OLDT, its depth is reduced from $U_{latt} \approx h \times 8$~MHz to $\gamma U_{latt}$ with $\gamma \lesssim 5 \times 10^{-3}$ [see Fig.~\ref{fig1}~(c)]. To source ion flows, we employ three-stage PI of the $5S_{1/2}$ atoms by the 795~nm and 762~nm lasers, which are also used for Rydberg-atom excitation, and the attenuated 1064-nm OLDT field [see Fig.~\ref{fig1}~(b)].  The ac light shifts of the bound atomic levels in the attenuated lattice of $< 100$~kHz are irrelevant during the 20-$\mu$s-long ionization and Rydberg-probe phase. With the PI cross section of $5D_{3/2}$ at 1064-nm of
$\approx 44$~Mb~\cite{cardman2021} and considering all three PI stages [see Fig.~1~(b)], the PI rate in the attenuated lattice is estimated to have a maximum value of $\lesssim 5 \times 10^{4}$~s$^{-1}$ per 5$S_{1/2}$ atom in our experiments. For up to $\sim 10^4$ atoms trapped in the OLDT mode, and assuming some inefficiency due to the beam profiles of the three PI laser beams and the atom distribution, the highest ion-source rates in our experiment are estimated $R_{ion} \lesssim 5 \times 10^8$~s$^{-1}$. Only at the highest ion-source rates may the atom sample become slightly depleted by PI-induced atom loss. PI of the Rydberg atoms used for electric-field measurement does not play a role.

\subsection{Electric-field calibration}

\begin{figure}[htb]
 \centering
  \includegraphics[width=0.48\textwidth]{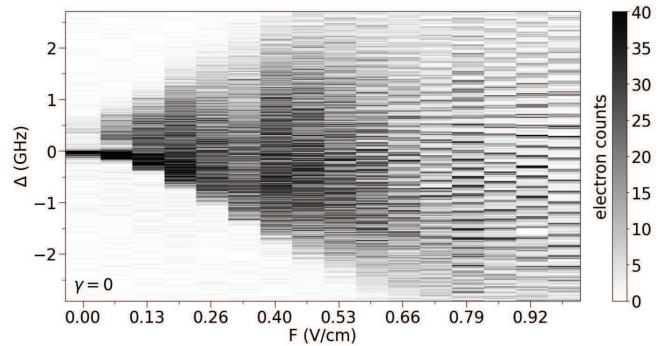}
  \caption{Stark map of 57$F_{5/2}$ and neighboring hydrogenic states versus applied field, $F$, and laser detuning, $\Delta$, in the absence of ions. The periodic set of resolved linear Stark states at $F \gtrsim 0.6$~V/cm is used to calibrate $F$.} 
  \label{fig2}
\end{figure}

\begin{figure*}[t!]
 \centering
  \includegraphics[width=\textwidth]{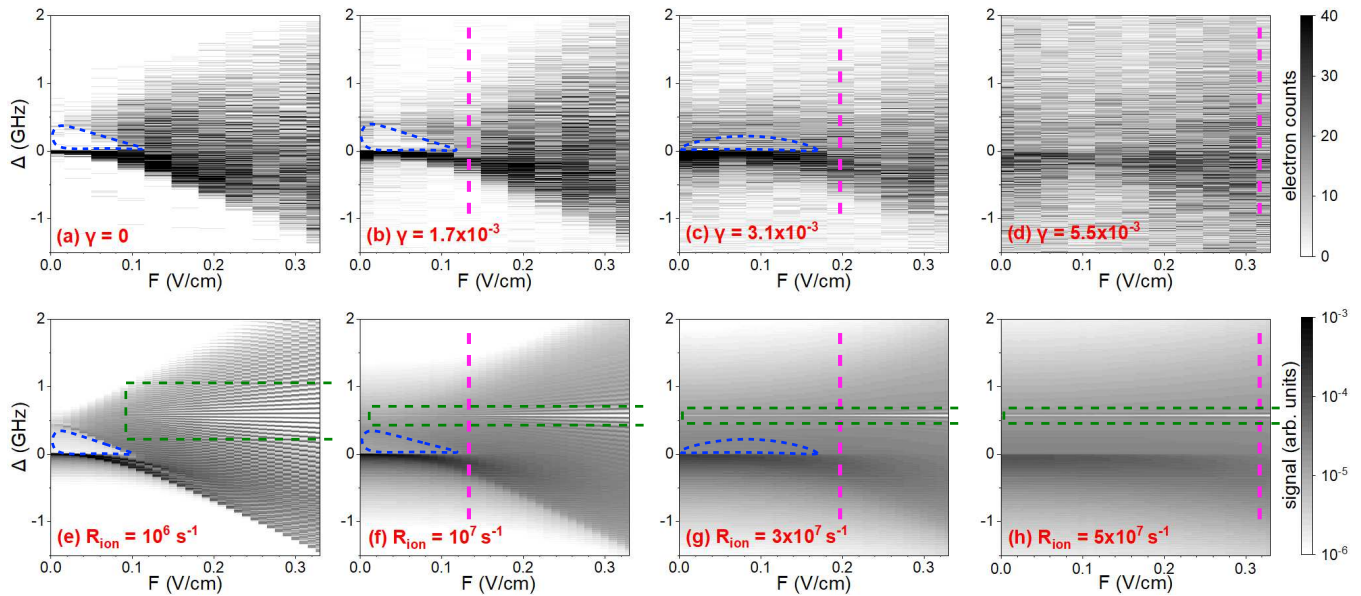}
  \caption{(a) - (d) Experimental Stark maps of 57$F_{5/2}$ and the neighboring hydrogenic states at the $\gamma$-values shown. 
  (e) - (h) Corresponding simulated Stark maps with empirically determined ion rates, $R_{ion}$, indicated on the plots.
  Dashed saucer-shaped areas in (a) - (c) and (e) - (g) indicate regions where the signal is small because state mixing and redistribution of oscillator strength from the $57F$ state to neighboring hydrogenic states is not complete. The dashed rectangles in (e) - (h) frame quasiperiodic structures that are present in the simulated results but absent in the experimental data. The short-dashed vertical bars indicate extraction fields $F$ at which the 57$F$ line becomes indiscernible in the signal due to mixing with hydrogenic states. See text for explanation and discussion.
 }
  \label{fig3}
\end{figure*}

To calibrate $F$, we acquire a Stark map without ions near the field-free $\ket{5D_{3/2}, F = 4} \rightarrow \ket{57 F_{5/2}}$ transition, shown in Fig.~\ref{fig2}. The $57F$ line at field $F=0$~V/cm
defines $\Delta=0$~MHz. In the field range $0 < F \lesssim 0.1$~V/cm, hydrogenic states with angular-momentum quantum number $\ell \ge 4$ begin to mix with $57F$ and generate signal at positive $\Delta$. As $F$ is increased further, the hydrogenic states spread out into both positive- and negative-$\Delta$ domains due to the linear Stark effect, and the signal increasingly spreads across all Stark states due to increased state mixing with the 57$F$ state. For $F \gtrsim$ 0.6~V/cm, the linear Stark states become well resolved and form a periodic structure of lines~\cite{Zimmerman.1979, li2012, peper2019}. Stark maps calculated for Rb Rydberg atoms with principal quantum number $n$ show a period of $3 n e a_0 F$ (not $1.5 n e a_0 F$, as in hydrogen). Matching calculated with measured spectra, the field $F$ is calibrated against applied voltage to an uncertainty of about $1\%$, which is satisfactory for the present purpose. In the presence of ion streams, the ionic macro- and microfields add to ${\bf{F}}$ according to Eq.~\ref{eq:e_net}. In the following sections, we use the resultant alterations in the Rydberg Stark spectra to diagnose the electric fields in the ion source.

\section{Results}
\label{sec:results}

We perform two sets of experiments, the first using the 
57$F_{5/2}$ state (Sec.~\ref{subsec:57F}), and the second using the 60$P_{1/2}$ state (Sec.~\ref{subsec:pstate}). Since the respective 
quantum defects are approximately $0.0165$ and $2.65$~\cite{gall}, 
these states are close in energy. In the first measurement we exploit the fact that the relevant Hilbert space has on the order of 100 near-degenerate 
states for each magnetic quantum number, $m_J$, that couple to each other in dc and in low-frequency ac fields, resulting in maximal linear Stark effect. 
Recently, such states have been utilized for detection of rf fields in VHF/UHF frequency ranges~\cite{brown2022}. In the second measurement, we use the fact that the two magnetic sublevels, $m_J = \pm 1/2$, of $60P_{1/2}$ have equal quadratic Stark shifts, allowing a straightforward extraction of electric-field distributions from spectroscopic line shapes. However, at $\sim 0.4$~V/cm, the highest field relevant in the present work, 
the quadratic Stark shift of $60P_{1/2}$ is still smaller than the maximal linear Stark shift for $n=57$ by a factor of $\sim 40$. Hence, the 60$P_{1/2}$-measurement loses accuracy at low fields.

\subsection{Experimental results and phenomenological interpretation of 57$F_{5/2}$ spectra in ion sources}
\label{subsec:57F}

In the top row of Fig.~\ref{fig3} we show experimental Stark maps near  57$F_{5/2}$ at four  $\gamma$-values, and in the bottom row a set of matching simulations discussed later. An initial examination of the top row of Fig.~\ref{fig3} reveals several features. 

(1) With increasing $\gamma$, the hydrogenic Stark states efficiently mix with the 57$F$ state, causing the overall spectral distribution to cover a region of several GHz, even at $F = 0$. At the largest $\gamma$, the spectral broadening at $F = 0$ exceeds 2~GHz. Taking into account that in a net electric field, $E_{net}$, the largest linear Stark shifts are $\approx \pm 1.5 n^2 e a_0 E_{net}$~\cite{gall}, one can conclude from the observed spectral broadening that the ion macro- and microfields, $\vert {\bf{E}}_{mac} + {\bf{E}}_{mic} \vert$, exceed a range of about $0.4$~V/cm at the highest source rates and at $F = 0$. Hence, the ionic fields exceed $F$ over a large fraction of the investigated range of $F$ and $\gamma$.

(2) At small $F$, the $57F$ signal exhibits a sharp dropoff near $\Delta = 0$~MHz toward positive $\Delta$. As $F$ increases, 
the $57F$ signal becomes diluted, red-shifts and eventually blends 
into the overall signal. This leads to a ``knee'' in the signals in Fig.~\ref{fig3} at certain $F$-values. With increasing $\gamma$, the knee dims and moves to larger values of $F$. To explain this behavior, we first note that fields $E_{net} \lesssim E_{crit} = 2 E_H \delta_f / (3 e a_0 n^5) $, with atomic energy unit $E_H \approx 27.2$~eV and $f$-quantum defect $\delta_f  \approx 0.0165$~\cite{han2006}, will not entirely mix the $F$ state into the manifold of hydrogenic states, resulting in a clearly visible 57$F$ signal near $\Delta = 0$~MHz that is slightly red-shifted due to a quadratic Stark effect~\cite{gall}. Whereas for fields $E_{net} \gtrsim E_{crit}$ the $57F$ state becomes mixed across many linear hydrogenic states, causing the 57$F$ signal near $\Delta = 0$~MHz to wash out and to become indiscernible from that of the hydrogenic states. Here, $E_{crit} \approx 0.1$~V/cm (as may also be inferred from Fig.~\ref{fig2}). The observed knee therefore occurs at extraction fields $F$ that are sufficiently strong that the net electric-field distribution,  $P(E_{net})$, has no significant contribution left in the range $E_{net} \lesssim E_{crit} = 0.1$~V/cm. The data in Fig.~\ref{fig3} show that, as $\gamma$ increases, it takes a larger $F$ to dilute the ion density to a degree at which the weak-field contribution, $\int_0^{\sim 0.1~{\rm{V/cm}}} P(E_{net}) dE_{net}$, becomes insignificant. At the highest ion rates in Fig.~\ref{fig3}, the ionic fields broaden the distribution $P(E_{net})$ by enough that even at $F = 0.35$~V/cm, the largest $F$ studied, a substantial low-field probability $E_{net} \lesssim E_{crit} = 0.1$~V/cm remains. This physical scenario is confirmed by the simulated field distributions $P(E_{net})$ shown below in Fig.~\ref{fig4}.

(3) Related to items (1) and (2), at low $\gamma$ it is observed that the combination of state mixing, Stark shifts and redistribution of oscillator strength from the 57$F$ into the hydrogenic states leaves a region of relatively small signal, indicated by the blue-dashed saucer-shaped areas in Fig.~\ref{fig3}. With increasing $\gamma$ the blue-dashed shapes increasingly fill in with signal and stretch to larger $F$, until they 
become unobservable above a certain $\gamma$. Following arguments given in item (2), as a threshold condition for the saucer-shaped areas in Fig.~\ref{fig3} to disappear we expect that $\gamma$ must be large enough that at $F = 0$ the peak of the ion-field distribution, $P(E_{net})$, exceeds $E_{crit} = 0.1$~V/cm.

(4) Comparing measurements and simulations in Fig.~\ref{fig3}, it is seen that the experiment does not resolve the quasiperiodic structures between the dashed green lines in the simulation. The latter are due to linear Stark states with near-zero slope, equivalent to near-zero electric-dipole moment (see Fig.~\ref{fig2}). This minor disagreement is likely caused by spectroscopic line broadening due to the MOT magnetic field (which is left on) and power broadening of the lower transitions driven by the 762-nm and 795-nm lasers. The structures between the dashed green lines in Fig.~\ref{fig3} could also be washed out by the inhomogeneity of the ion electric fields within the volume of individual Rydberg atoms, as one may expect from the spectra of individual Rydberg-atom-ion pairs~\cite{duspayev2021}.

\subsection{Model to synthesize simulated high-$\ell$ spectra}
\label{subsec:model}

To confirm our interpretation of the data presented in Sec.~\ref{subsec:57F}, we have developed a detailed model of the ion streams, their electric fields, and the Stark spectra that will result. 
In our ion-trajectory model, we assume an initial Gaussian atom density distribution with a full width at half maximum (FWHM) of 15~$\mu$m transverse to and 500~$\mu$m along the OLDT direction, an initial ion temperature of 44~mK, according to the ion recoil received in the PI, a fixed average ion rate $R_{ion}$, and a simulated duration of 20~$\mu$s. The ions are generated at random times between $t=0$ and $t=20~\mu$s and evolve under the influence of the ion extraction field, ${\bf{F}} = F\hat{\bf{x}}$, and the Coulomb fields of all ions. The ions give rise to both the macro- and microfields, ${\bf{E}}_{mac}$ and ${\bf{E}}_{mic}$. Every 500~ns the electric-field vectors are sampled on a three-dimensional grid of 1~$\mu$m spacing in all directions, inside a tube of 10~$\mu$m radius that is centered with the ion source. The sampling volume approximates the extent of the Rydberg-atom field-sampling volume. The simulation yields the net field distribution affecting the Rydberg atoms, $P(E_{net})$, with $E_{net} = |{\bf{F}}+  {\bf{E}}_{mac} + {\bf{E}}_{mic}|$ (see Eq.~\ref{eq:e_net}), as well as maps of the electric field averaged 
over user-defined time intervals, $\langle {\bf{E}}_{net} \rangle = {\bf{F}}+ \langle {\bf{E}}_{mac} \rangle $, maps of the root-mean-square (rms) electric field about the average, ion-density maps, as well as sample ion trajectories for visualization. A quasisteady-state is typically reached after 5~$\mu$s of ion sourcing. In the following, the distributions $P(E_{net})$ are averaged over the duration of the ionization and Rydberg field-probing pulse of $20~\mu$s, a time considerably longer than that needed for the system to reach a quasisteady-state. 

We also compute a bank of Stark spectra of Rydberg atoms in randomly polarized laser fields. The dc electric field applied to the atoms in the computation of the Stark maps is homogeneous and stepped in steps of 5~mV/cm from 0 to 1~V/cm. This results in 200 Stark spectra. We then use the electric-field  distribution $P(E_{net})$ from the trajectory simulation as a weighting function to obtain a weighted average of Stark spectra that models the experiment. The averaged Stark spectra depend on $F$ and the ion rate, $R_{ion}$, in the trajectory simulation. The $R_{ion}$-value is adjusted between simulation runs to arrive at a match between simulated and measured Stark spectra. The four simulation results shown in the bottom row of Fig.~\ref{fig3}
are obtained for the $R_{ion}$-values indicated in the figure.

The agreement observed in Fig.~\ref{fig3} between simulation and experiment is quite satisfactory. In results not shown, we have established that the exact volume of the field sampling region does not significantly affect the results, as long as the field sampling region does not exceed the ion-sourcing region. The quasiperiodic bars in the simulation at $\Delta \sim 500$~MHz, which are due to Stark lines with near-zero dipole moment, are absent from the experiment due to possible reasons explained in Sec.~\ref{subsec:57F} item (4). Further, the best-fitting $R_{ion}$ tends to scale as $(U_{lat} \gamma)^\kappa$ with a $\kappa > 1$.  Since $\gamma \lesssim 5 \times 10^{-3}$, the degree of decompression is fairly extreme, and $R_{ion}$ becomes reduced due to several effects. 
These include atom loss from the lattice, decompression-induced density reduction of the remaining trapped-atom cloud, and reduction in PI rate per atom (which is proportional to $\gamma$). The empirical finding $\kappa>1$ is therefore expected. For the experimental procedure employed to vary $R_{ion}$, it is only important that the function $R_{ion}(\gamma)$ is smooth and homogeneous, allowing control of $R_{ion}$ using $\gamma$ as a control parameter. At the present stage of the investigation, details of the function 
$R_{ion}(\gamma)$ are not critical.  

\subsection{Analysis and interpretation of simulated electric-field distributions}
\label{subsec:model2}

In the following, we discuss the physical picture that arises from the simulations that have successfully reproduced the experimental data in Fig.~\ref{fig3}.
The simulations provide us with a complete picture of the ion stream's particle phase space and electric-field distribution. Generally, we find in our simulations that the ion macrofields, $E_{mac}$, which are not to be confused with $F$, are considerably smaller than the microfields, $E_{mic}$. At $R_{ion} = 10^{8}$~s$^{-1}$ and $F=0$, $E_{mac}$ is less than 12$\%$ of the
rms-value of the microfields, while at $R_{ion} = 10^{7}$~s$^{-1}$ and $F=0$ it is less than about 5$\%$. Hence, the dominant ion-field effects are from the micro-fields.

\begin{figure}[htb]
 \centering
  \includegraphics[width=0.46\textwidth]{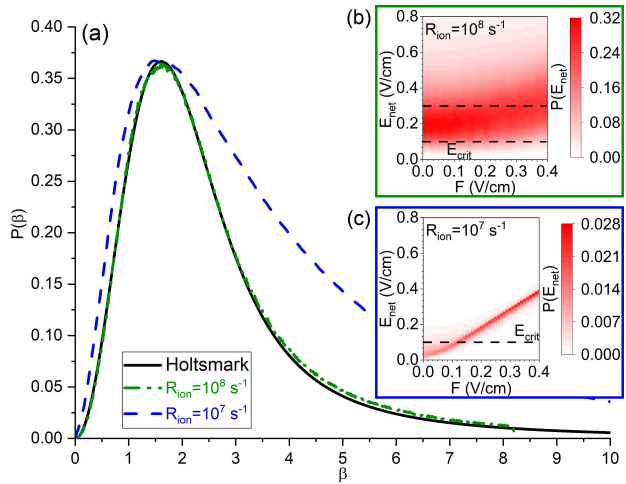}
  \caption{(a) Probability density function of scaled electric field, $\beta = E_{net}/\tilde{E}$ for the cases of a pure Holtsmark distribution (solid black line) and for ion sources with simulated ion rates of $R_{ion} = 10^8$~s$^{-1}$ (green dash-dotted line) and $R_{ion} = 10^7$~s$^{-1}$ (blue dashed line), for $F = 0$. The distributions are normalized to equal peak heights. 
  Panels (b) and (c) show probability density functions $P(E_{net})$ versus $E_{net}$ and $F$ for $R_{ion} = 10^8$~s$^{-1}$ and $R_{ion} = 10^7$~s$^{-1}$, respectively.}
  \label{fig4}
\end{figure}

First, we examine how close the electric-field distribution in the ion-source region can come to an ideal Holtsmark distribution, $\mathcal{P}_H (E_{net})$, for which ${\bf{F}} = {\bf{E}}_{mac} = 0$ in Eq.~\ref{eq:e_net}. The idealized microfield distribution in cases of negligible external and macrofield, $\mathcal{P}_H (E_{net})$, is given by the Holtsmark function~\cite{holtsmark}, which has been widely utilized in the context of plasma physics~\cite{hooper, potekhin, feldbaum, pohl2004, demura}, field sensing~\cite{osterwalder, anderson2017} and astrophysics~\cite{Chandrasekhar, Pietronero_2002}. The Holtsmark distribution has the form:
\begin{equation}
\mathcal{P}_H (E_{net}) = \frac{2}{\pi} \frac{E_{net}}{\tilde{E}^2} \int^\infty_0 dx~x \sin{\Big(\frac{E_{net}}{\tilde{E}} x \Big)} \exp{\Big(-x^{3/2}\Big)}.
\label{eq:Holstmark} 
\end{equation}
\noindent The electric-field scale parameter, $\tilde{E}$, is related to the ion density, $\rho_{ion}$, by
\begin{equation}
\tilde{E} = \frac{e}{2 \epsilon_0} \Big(\frac{4}{15} \rho_{ion} \Big)^{2/3}
\label{eq:mpEfield} 
\end{equation}
\noindent where $e$ is the electron charge and $\epsilon_0$ the vacuum electric permittivity. The function $\mathcal{P}_H (E_{net})$ peaks at  $E_{net} \approx 1.61 \tilde{E}$ and has an average field of about $2.9 \tilde{E}$. Due to the quadrupolar character of the electric field near random field zeros between ions, the low-field scaling behavior of $\mathcal{P}_H$ is $\propto E_{net}^2$.  

In Fig.~\ref{fig4}~(a) we compare the ideal Holtsmark distribution with  simulated electric-field distributions in the ion source, $P(E_{net}/\tilde{E}) = : P(\beta)$ for $F = 0$, with the scaling fields $\tilde{E}$ empirically adjusted to match the peaks of the simulated distributions with that of the Holtsmark distribution. It is seen that for $R_{ion} = 10^8$~s$^{-1}$, the largest $R_{ion}$ in Fig.~\ref{fig3}, the electric field follows a distribution that almost perfectly matches a Holtsmark distribution. Using the ion density $\rho_{ion}$ from the simulation, averaged over a cylinder of 10~$\mu$m radius and 200~$\mu$m length, from Eq.~\ref{eq:mpEfield} we compute a scaling field $\tilde{E}\approx$ 0.15~V/cm, whereas the empirical scaling field that yields the best match in Fig.~\ref{fig4}~(a) is $\approx$ 0.12~V/cm. We attribute the mild deviation to the finite extent of the ion distribution, which has a simulated FWHM diameter of the ion density of 25~$\mu$m and a Wigner-Seitz (WS) radius of 9.8~$\mu$m. Since the ratio between FWHM diameter and WS radius is only 2.5, a small deviation between the ideal and empirical scaling fields $\tilde{E}$ is expected.

In Fig.~\ref{fig4}~(a) we also show a result for $R_{ion} = 10^{7}$~s$^{-1}$, which is at the lower end of ion rates studied. In that case, the simulated scaled-electric-field distribution is nearly a factor of two wider than the Holtsmark distribution, which is attributed to fact that the WS radius ($17~\mu$m) is almost as large as the FWHM diameter of the ion density. Hence, the ion distribution tends to become one-dimensional along the OLDT axis, leading to the observed relative overabundance of stronger fields. Another consequence of the larger WS radius is that the empirical scaling field that yields the best peak match, displayed in Fig.~\ref{fig4}~(a), is only $\approx$~0.02~V/cm, while Eq.~\ref{eq:mpEfield} yields  $\tilde{E}\approx 0.05$~V/cm, for the simulated ion density at $R_{ion} = 10^{7}$~s$^{-1}$. The simulated field distribution at very low net fields still is $\propto E_{net}^2$. This is expected because the electric-field character near the accidental field zeros between ions remains quadrupolar, regardless of detailed parameters.

In Fig.~\ref{fig4}~(b) we display $P(E_{net})$ for $R_{ion} = 10^{8}$~s$^{-1}$ and  $F$ ranging from 0 to 0.4~V/cm. It is seen that
the response of the ion source to $F$ picks up very slowly, as $F$ is increased. Even at the largest $F$ there remains significant net probability for $E_{net} < E_{crit} = 0.1~$V/cm; also, there is substantial probability for fields $E_{net} \gtrsim  0.3~$V/cm across the entire range of $F$.  The features accord with the Stark spectra at the highest source rate in Fig.~\ref{fig3}, where the 57$F$ signal dropoff near $\Delta = 0$~MHz persists throughout the $F$-range, which is indicative of fields $E_{net} < E_{crit} =  0.1~$V/cm [see item (2) in Sec.~\ref{subsec:57F}]. Also, the Stark spectrum is broadened beyond a range of 2~GHz for all $F$, which is indicative of substantial probability for fields $E_{net} > 0.3~$V/cm [see item (1) in Sec.~\ref{subsec:57F}]. Finally, $P(E_{net})$ peaks well above 0.1~V/cm across the entire $F$-range, which is consistent with the absence of any saucer-shaped signal-depression regions in Figs.~\ref{fig3}~(d) and~(h) [see item (3) in Sec.~\ref{subsec:57F}].
It is noteworthy that at $F=0.35$~V/cm the distribution $P(E_{net})$ in Fig.~\ref{fig4}~(b) peaks visibly below $F$. This partial shielding of $F$  within the field-sensing volume is caused by a counterfield on the order of $0.1$~V/cm, pointing along $-\hat{\bf{x}}$, that the stream of extracted ions creates at the location of the Rydberg atoms. This counterfield is the most significant ionic macrofield effect seen in the present work.      

As seen in Fig.~\ref{fig4}~(c), at $R_{ion} = 10^{7}$~s$^{-1}$ the field distribution, measured on an absolute field scale, is much narrower than that at $R_{ion} = 10^{8}$~s$^{-1}$.
Due to its reduced field-shielding and Coulomb effects, the weak ion source responds quickly as $F$ is increased. As a consequence, the field distribution narrows down and peaks up near the value of $F$ already when $F \gtrsim 0.1$~V/cm [see Fig.~\ref{fig4}~(c)]. The absence of fields $E_{net} < 0.1~$V/cm in the range $F \gtrsim 0.15$~V/cm accords with the ``knee'' location observed in Fig.~\ref{fig3}~(b) [see item (2) in Sec.~\ref{subsec:57F}]. It is also seen that the peak of $P(E_{net})$ is well below 0.1~V/cm at $F=0$~V/cm and moves past 0.1~V/cm at $F \gtrsim 0.1$~V/cm. According to item (3) in Sec.~\ref{subsec:57F}, this causes the saucer-shaped regions of low signal in Figs.~\ref{fig3}~(b) and~(f) that stretch out to $ F \approx 0.1$~V/cm.

\subsection{Analysis using Rydberg $P_{1/2}$ states}

\label{subsec:pstate}

\begin{figure}[t]
 \centering
  \includegraphics[width=0.48\textwidth]{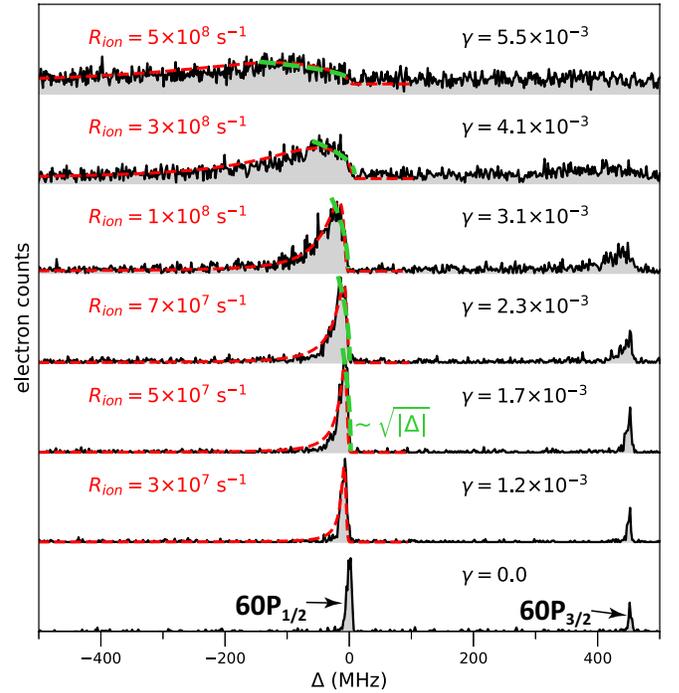}
  \caption{Black solid curves with gray-shaded area: experimental recordings of Stark-broadened and -shifted $60P_J$ lines at different $\gamma$-values. Red dashed: line profiles derived from simulated electric-field distributions for the indicated ion-source rates, $R_{ion}$. Green dashed: square-root functions that approximate the small-shift behavior of the spectral lines.}
  \label{fig5}
\end{figure}

In the following we describe results obtained from measurements using  $60P_J$ levels, which exhibit quadratic Stark shifts in weak electric fields. The timing of the OLDT amplitude reduction and the laser excitation is the same as in Fig.~\ref{fig1}~(c). We increase the power of the 762~nm laser for better signal-to-noise; the increased 762~nm-laser power also leads to larger values of $R_{ion}$. $60P_J$ spectra taken at a set of $\gamma$-values and $F=0$ are shown in Fig.~\ref{fig5}. The fine structure of the $60P$ state is well resolved. The laser detuning $\Delta$ is referenced to the transition $\ket{5D_{3/2}, F = 4} \rightarrow \ket{60 P_{1/2}}$ in the absence of ions. 
The stronger fine-structure component, $60P_{1/2}$, has a 
polarizability $\alpha=1122$~MHz/(V/cm)$^2$, while the polarizabilities of the $m_J$ components of $60P_{3/2}$ are 1353~MHz/(V/cm)$^2$ and 1139~MHz/(V/cm)$^2$. Denoting the level shift with $\Delta = - \alpha E^2 / 2$, an electric-field distribution $P(E_{net})$ maps into a distribution of 60$P_{1/2}$ shifts given by
\begin{equation} 
P_\Delta (\Delta) = P (E_{net}=\sqrt{|2 \Delta / \alpha|}) /\sqrt{|2 \Delta / \alpha|} \quad.
\label{eq:d1}
\end{equation}
For microfield distributions that scale as $E_{net}^2$ at small field, the distribution of shifts at small $\Delta$ scales as $P_{\Delta} \propto \sqrt{|\Delta|}$. The observed line shapes, $\tilde{P}_{\Delta}(\Delta)$, are given by the convolution of $P_{\Delta} (\Delta)$ with the field-free Rydberg line shape $g(\delta)$, which is taken to be a Gaussian with an FWHM of 5~MHz, according to the experimentally observed linewidths at $F = 0$ and low atom density,
\begin{equation} 
\tilde{P}_\Delta (\Delta) = \int P_\Delta(\Delta-\delta) g(\delta) d\delta \quad.
\label{eq:d2}
\end{equation}

As seen in Fig.~\ref{fig5}, near-symmetric initial spectra at $\gamma = 0$ evolve into highly asymmetric lines with extended red tails at increasing $\gamma$. The lower fine-structure component maintains sufficient signal to noise to allow for line-shape analysis over most of the $\gamma$-range studied. For $\gamma \gtrsim 2.3 \times 10^{-3}$ the line shift becomes large enough that the convolution with the line shape becomes a minor effect, and the electric-field distribution      
\begin{equation} 
P (E_{net}) \approx \tilde{P}_\Delta (\Delta = -\alpha E_{net}^2/2) |\alpha E_{net}| \quad,
\end{equation}
allowing, in principle, a direct extraction of the electric-field distribution from the experimental data. Some improvement at low $\gamma$ may be achieved by deconvolution of $\tilde{P}_\Delta (\Delta)$ using the known $g(\delta)$.
At $\gamma = 5.5\times 10^{-3}$ the 60$P_{1/2}$ component is broadened over several 100~MHz, and the 60$P_{3/2}$ component is barely visible above the background signal.

In the ion trajectory simulation, we empirically vary the ion rate $R_{ion}$, extract the field distributions $P(E_{net})$, and compute line profiles according to Eqs.~\ref{eq:d1} and~\ref{eq:d2}. $R_{ion}$ is varied until a good match is achieved. As seen in 
Fig.~\ref{fig5}, the empirical fits agree well with the experimental data, on an accuracy scale limited by experimental signal to noise.

A characteristic and easy-to-check discriminator of microfield-dominant distributions against distributions in approximately homogeneous fields is the low-field behavior, which scales as
$P(E_{net}) \sim E_{net}^2$ in microfields, whereas in distributions in approximately homogeneous fields there are no small fields. For lines with quadratic Stark shifts, microfields generate a line profile $\propto \sqrt{|\Delta|}$, whereas inhomogeneously broadened applied fields would shift the entire line and leave no signal near the original line position. In Fig.~\ref{fig5}, on the blue sides of the spectral lines we observe relatively sharp dropoffs that terminate at the field-free line position across the entire range of $\gamma$. This observation is consistent with line-shift distributions that scale  $\propto \sqrt{|\Delta|}$ at small $\Delta$. To emphasize the small-shift scaling, in Fig.~\ref{fig5} we have laid exact square-root functions over the blue sides of the field-shifted lines. Within the accuracy of our experiment, it is seen that the small-shift scalings follow a square-root behavior, indicating microfield dominance.

\section{Discussion}
\label{sec:disc}

\begin{figure}[htb]
 \centering
  \includegraphics[width=0.48\textwidth]{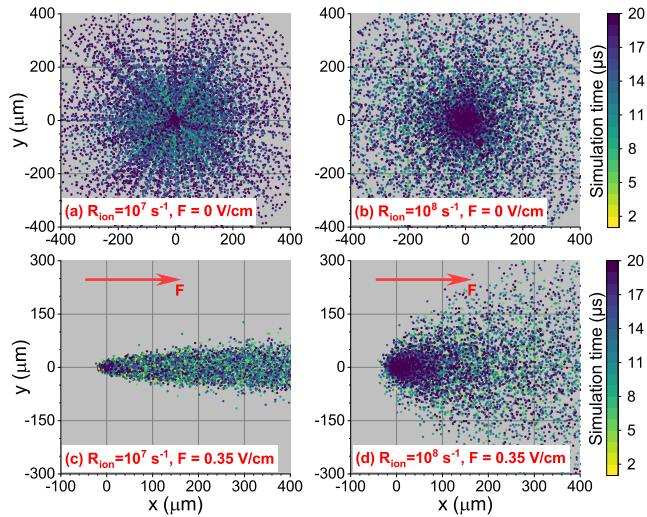}
  \caption{Ion trajectories projected into the $x-y$ plane for 20~$\mu$s of simulated dynamics and extraction fields $F = 0$ [(a), (b)] and $F = 0.35$~V/cm [(c), (d)], applied along the $x$-axis. The ion rates are $R_{ion} = 10^7$~s$^{-1}$ in (a), (c), and $10^8$~s$^{-1}$ in (b), (d).}
  \label{fig6}
\end{figure}

In the preceding sections we have established that Rydberg-atom-based electric-field sensing is well suited to acquiring information on the electric-field distribution in ion sources originating from PI of dense, laser-cooled atom clouds in optical dipole traps. It was also seen that in the highest-flux cases studied the electric-field distribution in the ion-sourcing region approaches that of an ideal Holtsmark distribution (see Fig.~\ref{fig4}). 

To further discuss the suitability of the presented setup to source and monitor ion streams, in the following we present direct images of ion streams at the lower and upper end of the experimentally covered flux regime, obtained from ion trajectory simulations at $R_{ion}=10^7~$s$^{-1}$ and $10^8~$s$^{-1}$, respectively, for extraction fields $F = 0$ and $F = 0.35$~V/cm. The images in the top row 
of Fig.~\ref{fig6} visualize that at $F = 0$ a larger $R_{ion}$ results in a more spread-out steady-state ion distribution than a low $R_{ion}$. More importantly, at $F = 0.35$~V/cm the directivity and spread of the extracted ion flow degrade considerably from low to high $R_{ion}$, as seen in the bottom row of Fig.~\ref{fig6}. 
Figs.~\ref{fig3} and~\ref{fig5} show that these cases result in qualitatively different spectral profiles of Rydberg-atom lines.
As an instance of a basic interpretation of the spectra, the vertical dashed bars in Fig.~\ref{fig3} mark the $R_{ion}$-dependent fields $F$ at which the 57$F$ line dissipates. At these fields the microfield dominance in the ion source wanes, the 
ion streams become directional, and the degradation due to microfields subsides. As such, it is seen that Rydberg-atom spectroscopy is well suited to characterizing ion streams derived from PI of localized cold-atom sources.

We finally consider how close the investigated quasisteady-state ion streams are to a plasma state, 
as this will have a bearing on how fluctuations in the ion sourcing might propagate through the ion streams in the form of waves. For an initial estimation, we have computed temperatures, $T_{ion}$, densities, $\rho_{ion}$, and Debye shielding lengths, $\lambda_D$, from averages taken over tubes of 10~$\mu$m radius and $200~\mu$m length, for $F = 0$ and $R_{ion}$ ranging from
$1 \times 10^{7}$~s$^{-1}$ to
$4 \times 10^{8}$~s$^{-1}$. We find that $T_{ion}$ ranges from 400~mK to 1.2~K, $\rho_{ion}$ from $4.7 \times 10^7$~cm$^{-3}$ to $6.4 \times 10^8$~cm$^{-3}$, and $\lambda_D$ from $6.3~\mu$m to $3~\mu$m.
Over the entire range, $T_{ion}$ substantially exceeds the initial ion temperature from PI of 44~mK due to continuous Coulomb effects. As expected, the temperature increases with $R_{ion}$. 
Over the $R_{ion}$-increase of a factor of 40, the 
ion density $\rho_{ion}$ increases only by about a factor of 14, which also reflects the increasing importance of continuous Coulomb expansion with increasing  $R_{ion}$. As a combination of the trends in  $T_{ion}$ 
and $\rho_{ion}$, the Debye length $\lambda_D$ drops by only about a factor of 2 over the entire $R_{ion}$-range. Comparing $\lambda_D$ with the FWHM diameter of the ion distribution of about 25~$\mu$m diameter, as $R_{ion}$ increases the system transitions from a marginally plasma-like state into a state that appears to be more solidly in the plasma regime. 

It follows that at the higher ion rates $R_{ion}$ studied in this work the system may exhibit plasma characteristics, such as a collective wave-like response in ion-flow behavior caused, for instance, by variations in the ion pump rate. This assessment is strengthened by estimates
of the ion sound speed, $v_s$, the ion plasma frequency, $f_{ion}$, and the ion response time for disorder-induced heating, $1/(4 f_{ion})$.
For $R_{ion}$ ranging from 
$1 \times 10^{7}$~s$^{-1}$ to
$4 \times 10^{8}$~s$^{-1}$ and $F = 0$, the respective values are estimated to range from 10 to 20~m/s, 150 to 500~kHz, and 1.6 to 0.5~$\mu$s. The heating time $1/(4 f_{ion})$ can be compared with the time it takes for an ion to traverse the 25-$\mu$m FWHM of the ion-density distribution, which is about 3~$\mu$s. As the system is in a dynamic steady-state flow, the spatial dependence of $v_s$, the ion-stream speed, $f_{ion}$, and so on will affect collective propagation phenomena, such as ion-acoustic waves and shock fronts through the ion streams. This topic is beyond the scope of the present work, but may be addressed in the future.

\section{Conclusion}
\label{sec:conc}

We have presented and analyzed Stark spectra of Rb $57F$ and $60P$ atoms in applied electric fields, and in macroscopic and microscopic electric fields within a cold-ion source. The source was implemented by PI of cold atoms concentrated within an optical dipole trap. The Rydberg electric-field sensing method described in this paper may work well for real-time, noninvasive monitoring of the performance of cold-ion sources in applications of focused ion beams (FIB) derived from cold atom clouds, mentioned in Sec.~\ref{sec:intro}. As such FIB sources already require a cold-atom source in a high-vacuum environment, the additional infrastructure needed to perform near-real-time, noninvasive ion-stream electric-field analysis along the lines described in the present paper would be fairly limited.  

The method also translates well to electric-field analysis in plasmas~\cite{feldbaum, park2010, anderson2017,Viray2020}. The analysis presented in our paper may be used to seek signatures of strong coupling in experimental microfield distributions.
In that context, it would be useful to increase the Coulomb coupling parameter,
$\Gamma = \frac{e^2 / (4\pi \epsilon_0 R_{WS})}{k_B T_{ion}}$, with WS radius $R_{WS}$ and ion temperature $T_{ion}$, from its present value of about 2 into a more deeply strongly coupled regime ($\Gamma \gg 1$).

\section*{ACKNOWLEDGMENTS}
We would like to thank Ryan Cardman, Dr. Jamie MacLennan and Dr. David Anderson for useful discussions. This material is based upon work supported by NSF Grant No. PHY-2110049, and by the U.S. Department of Energy, Office of Science, Office of Fusion Energy Sciences under award number DE-SC0023090.

\bibliography{references.bib}

\end{document}